\documentclass[prd
 ,twocolumn,a4paper
 ,secnumarabic%
,amssymb, amsmath,nobibnotes, aps,showpacs]{revtex4}
\usepackage{docs,bm,qtree}%
\expandafter\ifx\csname package@font\endcsname\relax\else
 \expandafter\expandafter
 \expandafter\usepackage
 \expandafter\expandafter
 \expandafter{\csname package@font\endcsname}%
\fi

\def\Journal#1#2#3#4{{#1} {\bf #2}, #3 (#4)}
\def\a{\alpha}
\def\b{\beta}
\def\g{\gamma}
\def\d{\delta}
\def\e{\epsilon}
\def\l{\lambda}
\def\s{\sigma}
\def\t{\theta}
\def\w{\omega}

\begin{document}

\title{Complete classification of purely magnetic, non-rotating\\
and non-accelerating perfect fluids.}

\author{Lode Wylleman}%
\email{lwyllema@cage.ugent.be}
\author{Norbert \surname{Van den Bergh}}
\email{norbert.vandenbergh@ugent.be}
\affiliation{Faculty of Applied Sciences TW16, Gent University, Galglaan 2, 9000 Gent, Belgium}
\date{\today}
\begin{abstract}
Recently the class of purely magnetic non-rotating dust spacetimes
has been shown to be empty (Wylleman, Class.\ Quant.\ Grav.\ 23, 2727). It turns out that purely
magnetic rotating dust models are subject to severe integrability
conditions as well. One of the consequences of the present paper is
that also rotating dust cannot be purely magnetic when it is of
Petrov type D or when it has a vanishing spatial gradient of the energy density.
For purely magnetic and non-rotating perfect fluids on the other
hand, which have been fully classified earlier for Petrov type
D (Lozanovski, Class.\ Quant.\ Grav.\ 19, 6377), the fluid is shown to be non-accelerating if
and only if the spatial density gradient vanishes. Under these
conditions, a new and algebraically general solution is found, which
is unique up to a constant rescaling, which is spatially homogeneous
of Bianchi type $VI_0$, has degenerate shear and is of Petrov type I($M^\infty)$ in the extended Arianrhod-McIntosh classification.

The metric and the equation of state are explicitly constructed and properties of the model are briefly discussed. We finally situate it
within the class of normal geodesic flows with degenerate shear tensor.\\
\end{abstract}
\maketitle

\section{Introduction}

For a given spacetime geometry,
the electric and magnetic parts of the Weyl tensor $C_{abcd}$
w.r.t.~some unit timelike congruence $u^a$
are pointwise defined by
\begin{eqnarray}\label{EH}
    E_{ab} &=& C_{acbd}\,u^c\,u^d,\\
    H_{ab}&=& \frac{1}{2}{\eta_{ac}}^{mn}C_{mnbd}\,u^c\,u^d,
\end{eqnarray}
$\eta_{abcd}$ being the spacetime permutation tensor.
$E_{ab}$ and $H_{ab}$ are traceless and symmetric tensors satisfying
$H_{ab}u^b=E_{ab}u^b=0$, and determine the Weyl tensor
completely~\cite{Matte,Bel,Ferrando1}. They were first introduced
(for the vacuum Riemann tensor) by Matte~\cite{Matte} when searching
for gravitational quantities playing an analogous role to the
electric and magnetic field in classical electromagnetism. Using the
decomposition (\ref{EH}), the Bianchi identities take a form
analogous to Maxwell's equations  for the electromagnetic
field~\cite{MaBa}. A non-conformally flat spacetime for which
$E_{ab}$, resp.~$H_{ab}$, vanish identically w.r.t.~some ${u_0}^a$
has therefore been called \emph{purely magnetic} (PM),
resp.~\emph{purely electric} (PE), and its Weyl tensor is said to be
PM, resp.~PE, w.r.t.~${u_0}^a$. As $E_{ab}$ and $H_{ab}$ (w.r.t.~any
$u^a$) are diagonalizable tensors, the Petrov type of PE or PM
spacetimes is necessarily I or D, and in each point ${u_0}^a$ is a
Weyl principal vector (which is essentially unique for Petrov type
I, and which is an arbitrary timelike vector in the plane of
repeated principal null directions for Petrov type D)
\cite{Barnes,Lozanovski5}.
The PM and PE property may be characterized independently of $u^a$,
as follows.
Defining
\begin{eqnarray}\label{Qab}
    Q_{ab}&=& E_{ab}+iH_{ab},
\end{eqnarray}
the
quadratic, cubic and 0-dimensional invariants $I,J$ and $M$ of the
Weyl tensor~\cite{McIntosh2,Penrose} can be written as
\cite{Matte,McIntosh1}
\begin{eqnarray}
    I&=&{Q^a}_b {Q^b}_a\nonumber\\
    &=& \l_1^2+\l_2^2+\l_3^2=-2(\l_1\l_2+\l_2\l_3+\l_3\l_1)\nonumber \\
    &=&({E^a}_b {E^b}_a-{H^a}_b {H^b}_a)+2i{E^a}_b {H^b}_a,\\
    J&=&{Q^a}_b {Q^b}_c {Q^c}_a\nonumber \\
    &=& \l_1^3+\l_2^3+\l_3^3=3\l_1\l_2\l_3 \nonumber \\
    &=& {E^a}_b {E^b}_c {E^c}_a-3{E^a}_b {H^b}_c {H^c}_a\nonumber\\
    & & -i({H^a}_b {H^b}_c {H^c}_a-3{E^a}_b {E^b}_c {H^c}_a), \label{eq:J}\\
    M&=&\frac{I^3}{J^2}-6\nonumber\\
    &=&\frac{2(\l_1-\l_2)^2(\l_2-\l_3)^2(\l_3-\l_1)^2}{9\l_1^2\l_2^2\l_3^2}, \label{eq:M}
\end{eqnarray}
where the solutions $\l_i$ of $\l^3-\frac{1}{2}\l I -\frac{1}{3}J=0$
are the non-zero eigenvalues of $Q_{ab}$ for any $u^a$. Hence, a
spacetime is PE (PM) if and only if all $\l_i$ are real (imaginary),
or~\cite{McIntosh1}
if and only if $M$ is real non-negative or infinite and $I$ is real
positive (negative). For PE or PM spacetimes, $M=0$ if and only if
the Petrov type is $D$; the respective types in the algebraically
general case were logically denoted $I(M+)$ and $I(M^\infty)$ in
the extended Petrov classification of Arianrhod and
McIntosh~\cite{Arianrhod1}, where $I(M^\infty)$ corresponds to
$J=0$, i.e.~to one of the $\l_i$ being identically zero.

Whereas large and physically important classes of examples exist for
PE spacetimes (for example all the static spacetimes are purely
electric\cite{Kramer}), only a few examples exist of the purely
magnetic ones. This is particularly true for the vacuum solutions
(with or without $\Lambda$ term), where no PM solutions are known at
all. This has lead to the conjecture that PM vacua do not
exist~\cite{McIntosh1}, but so far this has only been proved for
Petrov types D~\cite{McIntosh1} and $I(M^\infty)$~\cite{Brans}
or when the timelike congruence $u^a$ is shear-free~\cite{Haddow},
non-rotating~\cite{Trumper,VdB1}, geodesic~\cite{VdB2}, or satisfies
certain technical generalizations of these conditions
\cite{Ferrando2,Zakhary1}. In \cite{Ferrando3,Ferrando4} the
non-existence of shear-free or non-rotating PM models was
generalized to spacetimes with a vanishing Cotton tensor. As a
positive example on the other hand, the metric constructed in
\cite{Arianrhod2} turns out to be a PM kinematic counterpart to the
G\"{o}del metric~\cite{Maartens2}, but its source is unphysical as
the Ricci tensor is of Segr\'{e} type $[11,Z\overline{Z}]$. In
\cite{Lozanovski3}, PM locally rotationally symmetric (LRS)
spacetimes were shown to belong to either class I or III of the
Stewart-Ellis classification~\cite{StewartEllis},
the possible Segr\'{e}-types were determined and the most
general metric forms were
found, exhibiting one arbitrary function and three parameters.\\

In a cosmological context, perfect fluid models are studied, the
metric $g_{ab}$ being a solution of the Einstein field equation with
perfect fluid source term
\begin{eqnarray}\label{fieldeq0}
    R_{ab}-\frac{1}{2}R\,g_{ab}=(\mu+p)u_a\,u_b+p g_{ab}.
\end{eqnarray}
Here
$R_{ab}$ is the Ricci tensor, which is assumed not to be of
$\Lambda$-type ($\mu+p\neq 0$). In this case, the average 4-velocity
field $u^a\equiv {u_{\textrm{pf}}}^a$ of the fluid plays the role of a
geometrically preferred unit timelike vector field on the spacetime.
A possible cosmological constant $\Lambda$ has been absorbed in the
fluid's energy density and pressure by means of re-defining
$\mu=\mu^\prime+\Lambda$ and $p=p^\prime-\Lambda$.
The electric and magnetic parts (\ref{EH}) w.r.t.~${u_{\textrm{pf}}}^a$ then
represent the locally free gravitational field (not pointwise
determined by the matter content via the Einstein equations
(\ref{fieldeq0})), while the vanishing of their spatial divergence
together with the non-vanishing of the so called `curl' and
`distortion' parts of their spatial derivatives form necessary
conditions for the existence of gravitational wave perturbations of
the homogeneous and isotropic FRW spacetime
\cite{Hawking,Bruni,Dunsby,Maartens}. Whereas the electric part
$E_{ab}$ is the general relativistic generalization of the tidal
tensor in Newtonian theory~\cite{EllisDunsby}, the magnetic part
$H_{ab}$ has no Newtonian analogue and it's role is still poorly
understood (see however \cite{Lozanovski1,Bonnor,Baez}.

This motivates the attempt to construct purely magnetic perfect
fluid models with ${u_0}^a={u_{\textrm{pf}}}^a$ in the above. A
solution to (\ref{fieldeq0}) of this kind will then be called a
PMpf. The only known perfect fluid which is purely magnetic in a
certain region of spacetime, but which is not a PMpf in the above sense
(${u_0}^a\neq{u_{\textrm{pf}}}^a$), is the Lozanovoski-McIntosh metric \cite{LozMcInt}.

The present study is restricted to PMpf models, with
the exception of section 4.4. First however we conclude this introduction with an overview of all performed studies
and known examples of PMpf's, to the best of our knowledge.

In the case where the Petrov type is D, an exhaustive classification of PMpf's has by now been obtained, essentially due to two results. Firstly, it was shown most recently~\cite{VdBWyll} that Petrov
type D PMpf's are necessarily LRS class I or III, whereas secondly the most general metric forms~\footnote{However, for LRS class I a complicated third order ODE needs to be solved. For LRS class III the metric was shown to be explicitly constructable, but a simpler form than the one suggested in \cite{Lozanovski3} can be obtained by directly taking the function $s(t)$ as a new time-variable $u$ instead of $t$; the result is
$C^2 ds^2 = \frac{1}{\Gamma} \left(-\frac{1}{\Gamma}du^2 + u^2 \times \right.$ $\left. [dx+qe^B(zdy-ydz)]^2+e^{2B}(dy^2+dz^2)\right)$,
where $C, c$ and $q$ are constants and $k$ equals -1, 0 or 1, $B=-\log{(1+\frac{k}{4}(y^2+z^2))}$ and $\Gamma=2 q^2 u^4-k u^2+c$.} for each of these classes were derived in \cite{Lozanovski3}. Four systematic studies historically precede these results: on the one hand LRS space-times in general were
reinvestigated in \cite{vanElst1,Marklund} by two different methods,
hereby also briefly discussing the purely magnetic case; on the other hand shear-free~\cite{Lozanovski5}, resp.\ non-rotating~\cite{Lozanovski2} PMpf's of Petrov type D were shown to be LRS class I, resp.\ III. Finally the following  three  distinct PMpf solutions are particular cases of the general LRS cases above: the
$p=\frac{1}{5}\mu$ Collins-Stewart space-time~\cite{Collins} and
Lozanovski-Aarons metric~\cite{Lozanovski1}, both
LRS class III, and the LRS class I rigidly rotating axistationary
model with circular motion of \cite{Fodor}.
The last example was found within a study (regardless of the Petrov type) of axistationary perfect fluids, which concentrated mostly on the PMpf subclass. Concretely, it was shown that axistationary PMpf's
with circular motion necessarily have non-vanishing vorticity and spatial-3 gradient of the matter density, and that such spacetimes are LRS class I.\\

No algebraically general PMpf's have been
found yet, although they might be of relevance for
cosmological modelling. The PMpf subclass of irrotational dust
spacetimes consists of `silent' universe models
\cite{Matarrese2,vanElst2} and was investigated in \cite{Maartens2},
but the appearance of chains of severe integrability conditions
(analogously as for the widely studied purely electric subclass in
the Petrov type I case~\cite{vanElst2,Sopuerta}) made the authors
conjecture that this subclass might be empty. This was recently
proved in \cite{Wylleman} for the case of dust, regardless of Petrov
type and cosmological constant.

The present study continues this line
of investigation. Whereas evidence is provided that the class of PM
rotating dust must be severely restricted as well,  we present as a
main result a first example of an algebraically general PMpf
solution, which is both non-rotating and non-accelerating. Up to a
constant rescaling, this solution is shown to be the unique PMpf
with these properties. Moreover it is spatially homogeneous with
degenerate shear tensor and, just as the related LRS class III
models of \cite{Lozanovski2}, it turns out to satisfy the energy
conditions in an open subset of spacetime.

The structure of the paper is as follows. In section 2 we set up the
basic variables and equations for PMpf's, in a mixed 1+3
covariant/orthonormal tetrad approach. After pointing out why
solution families of the PM rotating dust class should be rather
poor in number, the subclasses of Petrov type D and of vanishing
spatial 3-gradient of the energy density are shown to be empty in
section 3. Section 4 provides a characterization of the new metric,
discusses its mathematical and physical features and situates it
within the broader context of \cite{VdB3}. We end with a conclusion
and discussion in section 5.

\section{Basic equations for PMpf's}


We use units such that $8\pi G=c=1$, Einstein summation convention,
the signature $(-,+,+,+)$ for spacetime metrics  $g_{ab}$, and
abstract (Latin) index notation~\cite{Wald} for tensorial quantities
(except for basis vector fields which are written in bold face
notation); round (square) brackets around indices denote (anti-)symmetrization. 
The perfect fluid field equations (\ref{fieldeq0}) may be rewritten as
\begin{eqnarray}\label{fieldeq}
        R_{ab}=\frac{1}{2}(\mu+3p)u_a\,u_b+\frac{1}{2}(\mu-p)h_{ab}.
\end{eqnarray}
Here ${h}_{ab}={g}_{ab}+u_a u_b$ projects orthogonally to the
fluid's 4-velocity field $u^a$, `spatializing' indices of tensorial
quantities by contraction; if this contraction is the identity
operation for all indices, the tensor is called spatial. For
covariant differential operations orthogonal to $u^a$,
the streamlined notation of \cite{MaBa,Maartens2,Maartens1} is the
most transparent. The covariant spatial derivative $D$ and the
associated curl and divergence (div) operators, acting on vectors
and 2-tensors, are defined as:
\begin{eqnarray}
\label{defD}& & D_a{S^{c\ldots d}}_{e\ldots f} = {h_a}^b{h^c}_p\cdots {h^{d}}_{q}{h_{e}}^{r}\cdots {h_{f}}^{s} \nabla_b{S^{p\ldots q}}_{r\ldots s},\\
    \label{divcurlV}& & \textrm{div}\, V = D_a V^a,\quad{\textrm{curl}\, V}_a=\e_{abc}D^b\,V^c.\\
  \label{divcurlS}& & {\textrm{div}\, S}_a = D^b S_{ab},\quad{\textrm{curl}\, S}_{ab}=\e_{cd(a}D^c {S_{b)}}^d.
\end{eqnarray}
Here $\nabla_c$ is the covariant derivative associated with the
Levi-Civita connection and $\e_{abc}=\eta_{abcd}u^d$ is the spatial
permutation tensor. The kinematics of the perfect fluid
$u^a$-congruence are then described~\cite{Ellis} by its acceleration
$\dot{u}_a$, vorticity $\omega_a=-\frac{1}{2}{\textrm{curl}\,u}_a$
and expansion tensor $\t_{ab}=D_{(a} u_{b)}$, which are all spatial.
Here and in general, a dot denotes covariant (`time')
differentiation along $u^a$. For scalar functions $f$ and vectors
$v^a$ one has~\cite{MaBa,Maartens1}
\begin{eqnarray}
 (D_a f)^\cdot  &=& D_a \dot{f}-{\t_{a}}^bD_b f-\e_{abc}\w^b D^c f   \nonumber \\
 \label{comTD} & & + \dot{u}_a \dot{f} +u_a\dot{u}^b D_b f, \\
\label{Riccif} \textrm{curl}(Df)_a &=& -\dot{f}\w_a,\\
\label{curlfv} \textrm{curl}(f\,v)_a &=& f\textrm{curl}\,v_a+\e_{abc}D^b f v^c.
\end{eqnarray}
The expansion tensor is further decomposed as
$\t_{ab}=\s_{ab}+\frac{1}{3}\t h_{ab}$, with $\s_{ab}=D_{(a}
u_{b)}-\frac{1}{3}\theta h_{ab}\equiv Du_{\langle ab\rangle}$ the
trace-free shear tensor of the fluid and
$\theta={\t^a}_a=\textrm{div}\,u$ its scalar expansion rate. In
general, $S_{\langle
ab\rangle}\equiv{h_a}^c{h_b}^dS_{(cd)}-\frac{1}{3}S_{cd}h^{cd}h_{ab}$ is
the spatially projected, symmetric and trace-free part of $S_{ab}$,
while $V_{\langle a\rangle}={h_a}^bV_b$ denotes the spatial
projection of $V_a$~\cite{Maartens}.

In a 1+3 covariant `threading' approach, the tensorial quantities
$\dot{u}_a,\omega_a,\s_{ab},\t$
and $E_{ab},H_{ab},\mu,p$
are taken as the fundamental dynamical fields. One focusses on the
Ricci-identity for $u_a$ and the Bianchi-identities, wherein the
Ricci-tensor is substituted for the right hand side of
(\ref{fieldeq}). They can be split in
constraint equations (involving only spatial derivatives) and
propagation equations. For a general perfect fluid, the
Ricci-equations become, in our notation:
\begin{eqnarray}
    \label{divs} & & {\textrm{div}\,\s}_a  - \frac{2}{3}D_a\t-{\textrm{curl}\,\w}_a + 2\e_{abc}\w^b\dot{u}^c = 0,\\
    \label{divw} & & \textrm{div}\,\w - \dot{u}^a\omega_a = 0,\\
    \label{defH} 
     & & {\textrm{curl}\,\s}_{ab}+D_{(a}\w_{b)} - H_{ab}+ 2\dot{u}_{\langle a}\omega_{b\rangle} = 0,\\
 & & \dot{\s}_{\langle ab\rangle} + \frac{2}{3}\theta\,\s_{ab}+ \s_{c\langle a}{\s_{b\rangle}}^c+
    \omega_{\langle a}\omega_{b\rangle} \nonumber \\
\label{dotsigma} & & - D_{\langle a}\dot{u}_{b\rangle} - \dot{u}_{\langle a}\dot{u}_{b\rangle}+ E_{ab}= 0,\\
     & & \dot{\theta}+\frac{1}{3}\theta^2+\s_{ab}\s^{ab}-2\omega^a\omega_a-\textrm{div}\,
    \dot{u}\nonumber \\
    \label{dottheta} & & -\dot{u}_a\dot{u}^a+\frac{1}{2}(\mu+3p) = 0,\\
    \label{dotw}& & \dot{\w}_{\langle a \rangle}+ \frac{2}{3}\t\w_a- \s_{ab}\w^b+\frac{1}{2}\textrm{curl}\,\dot{u}_a = 0.
\end{eqnarray}
The equations of conservation of momentum and energy (contracted Bianchi-identities) are:
\begin{eqnarray}
  \label{Dp} D_a p &=& -(\mu+p)\dot{u}_a,\\
  \label{dotmu} \dot{\mu} &=& -(\mu+p)\t.
\end{eqnarray}
For PMpf's $E_{ab}=0$, the remaining Bianchi-identities are
\begin{eqnarray}
  \label{divE} 
    & & [\s,H]_a-3H_{ab}\w^b+\frac{1}{3}D_a\rho=0,\\
    \label{divH} & & {\textrm{div}\,H}_a -(\mu+p)\w_a = 0,\\
     \label{dotE} & & {\textrm{curl}\,H}_{ab}- \frac{1}{2}(\mu+p)\s_{ab}+2\dot{u}^c\e_{cd(a}{H_{b)}}^d = 0,\\
     \label{dotH} & & \dot{H}_{\langle ab\rangle}+ \theta H_{ab}- 3\s_{c\langle a}{H_{b\rangle}}^c + \omega^c\e_{cd(a}{H_{b)}}^d = 0.
\end{eqnarray}
Here $[S,T]_a\equiv\e_{abc}S^{bd}{T_d}^c$ is the vector dual to the commutator of spatial tensors $S_{ab}$ and $T_{ab}$.
Note that in general, the term $E_{ab}$ in (\ref{dotsigma}) couples the evolution of the kinematical quantities to that of the Weyl tensor,
which is no longer the case for PMpf's. The constraints (\ref{divE}) and (\ref{dotE}) express ${\textrm{div}\,E}_a=0$ and $\dot{E}_{ab}=0$, respectively.

In an orthonormal tetrad approach, a specific orthonormal basis of
vector fields ${\cal B}=\{{{\mathbf e}_0}={\mathbf u},{\mathbf
e}_{\a}\}$ is taken. Here and below, Greek indices run from 1 to 3,
expressions containing these have to be read modulo 3
(e.g.~$X_{\a+1\a-1}=X_{12}$ for $\a=3$) and $\partial_\a f$ is
written for the action of ${\bf e}_\a$ on functions $f$. For spatial
tensorial quantities, only the components with Greek indices survive; one has in particular $h_{\a\b}=\delta_{\a\b}$, 
and $\e_{\a\b\g}$ becomes the 
alternating symbol on three indices, where we take the 
convention~\footnote{Following
\cite{Ellis}, different from the convention in
\cite{Kramer}, where $\eta_{1230}=-1$ is taken.} $\e_{123}=\eta_{1230}=1$. The commutator
coefficients ${\g^a}_{bc}=-{\g^a}_{cb}$ and Ricci-rotation
coefficients ${\Gamma^a}_{bc}$ of ${\cal B}$ are defined by
\begin{eqnarray}
    \label{commrel} [{\bf e}_b,{\bf e}_c]={\g^a}_{bc}{\bf e}_a,\quad \nabla_c {\bf e}_b={\Gamma^a}_{bc}{\bf e}_a.
\end{eqnarray}
As for any rigid frame, the lowered coefficients
${\Gamma}_{abc}=g_{ad}{\Gamma^d}_{bc}=-\Gamma_{bac}$ and
${\g}_{abc}=g_{ad}{\g^d}_{bc}$ are one-one related by
\begin{eqnarray}
    {\gamma}_{abc}=-2{\Gamma}_{a[bc]},\quad {\Gamma}_{abc}=
    \frac{1}{2}(\g_{bac}+\g_{cab}-{\g}_{abc}).
\end{eqnarray}
Combinations hereof, together with $\mu,p$ and the components $H_{\a\b}$
play the role of basic variables. In the present paper, we will
use\footnote{The $\Omega_\a$ are the non-zero components of the
angular velocity vector of the triad $\{{\bf e}_\a\}$ w.r.t.~the
`inertial compass', see \cite{MacCallum} and references therein. The
notation $n_{\a\a}$ is also in agreement with~\cite{MacCallum},
where the further decomposition
${\gamma^\a}_{\b\zeta}=\epsilon_{\b\zeta\kappa}n^{\kappa\a}+\delta^\a_\zeta
a_\b -\delta^\a_\b a_\zeta$, $n^{\a\b}=n^{(\a\b)}$ of the purely
spatial coefficients is exploited.}

\begin{eqnarray}
 \label{myhtheta}  h_\a &\equiv& \Gamma_{\a+1\,0\,\a+1}-\Gamma_{\a-1\,0\,\a-1} \\
&=& \s_{\a+1\,\a+1}-\s_{\a-1\,\a-1}\nonumber  \\
 \theta &=& {\Gamma^\b}_{0\b}\quad\textrm{or } \nonumber \\
 \label{thetaa} \t_{\a\a} &=& \Gamma_{\a0\a}=\s_{\a\a}+\frac{1}{3}\t,\\
     \s_{\a+1\,\a-1} &=& \frac{1}{2}(\Gamma_{\a+1\,0\,\a-1}+\Gamma_{\a-1\,0\,\a+1}),\\
 \label{eqrefomega}  \w_\a &=& \frac{1}{2}(\Gamma_{\a+1\,0\,\a-1}-\Gamma_{\a-1\,0\,\a+1}),\\
   \dot{u}_\a &=& \Gamma_{\a 00},\\
   \Omega_\a &=& {\Gamma}_{\a-1\,\a+1\, 0},\\
   \label{qa} q_\a &\equiv & {\gamma}_{\a-1\,\a-1\,\a}=-{\Gamma}_{\a\,\a-1\,\a-1},\quad \nonumber \\
   r_\a &\equiv & {\gamma}_{\a+1\,\a\,\a+1}={\Gamma}_{\a\,\a+1\,\a+1},\\
    n_{\a}&\equiv & {\Gamma}_{\a+1\,\a-1\,\a}\quad\textrm{or }\nonumber \\
    \label{naa} n_{\a\a} &=& \gamma_{\a\,\a+1\,\a-1}=n_{\a+1}+n_{\a-1}.
\end{eqnarray}

In the line of the Cartan one-form formalism, the basic equations
are the commutator relations (\ref{commrel}), the components of the
Ricci-identities and of the (second) Bianchi-identities
(\ref{Dp})-(\ref{dotH}) (see e.g.~\cite{Kramer} $\S 7$). The
Ricci-identities may be further split into\\
(a) the Jacobi-identities
(first Bianchi-identities)
\begin{eqnarray}
    \frac{1}{6} \begin{pmatrix} &d&\\a&b&c\end{pmatrix} & \equiv & 
    [e_{[a},[e_b,e_{c]}]]^d \nonumber \\
    &=& \partial_{[a}{\gamma^d}_{bc]}+{\gamma^f}_{[bc}{\gamma}^d_{a]f}=0,
\end{eqnarray}
with $\begin{pmatrix} &0&\\1&2&3\end{pmatrix}$ = (\ref{divw}) and $\begin{pmatrix} &0&\\0&\a+1&\a-1\end{pmatrix}$
 = (\ref{dotw}),\\
(b) the defining equations (\ref{defH}) and
(\ref{dotsigma}) for $H_{\a\b}$ and $E_{\a\b}$, and \\
(c) the components of the Einstein equations
(\ref{fieldeq}), where the $(00)$ component is Raychaudhuri's equation
(\ref{dottheta}) and the $(0\a)$ component is (\ref{divs}), whereas the $(\a\b)$ component
(which is not covered by the Ricci-identity for $u_a$) reads
\begin{eqnarray}
\partial_c{\Gamma^c}_{(\a\b)}&-&\partial_{(\b}{\Gamma^c}_{\a) c}+
{\Gamma^c}_{dc}{\Gamma^d}_{(\a\b)}-{\Gamma^c}_{\a f}{\Gamma^f}_{\b c}\nonumber \\
&=& \frac{1}{2}(\mu-p)\delta_{\a\b}.
\end{eqnarray}
Note that (\ref{dottheta})-(\ref{divE}) are equalities between
spatial tensorial quantities, which are moreover symmetric and
trace-free in the case where they are 2-tensors. The non-trivial
components are readily deduced from (\ref{defD})-(\ref{divcurlS}), (\ref{qa})-(\ref{naa})
and
\begin{eqnarray}
    D_\a V_\b &=& \partial_\a V_\b-V_\d\,\Gamma^\d_{\b\a},\nonumber \\
    \dot{V}_\a &=& \partial_0 V_\a + \e_{\a\b\g}\Omega^\b V^\g,\\
    D_\a S_{\b\g} &=& \partial_\a S_{\b\g}-S_{\d(\g}\Gamma^\d_{\b)\a}, \nonumber \\
    \dot{S}_{\a\b} &=& \partial_0 S_{\a\b} + 2\e_{\g\d(\a}\Omega^\g {S_{\b)}}^\d.
\end{eqnarray}

\section{No-go results for PM rotating dust}

The identity (\ref{Riccif}) applied to $f=p$, together with (\ref{Dp}) and (\ref{curlfv}), yields
\begin{eqnarray}\label{Ricciu}
    (\mu+p)\,\textrm{curl}\,\dot{u}_a+\e_{abc}D^b\mu\,\dot{u}^c=-2\dot{p}\w_a.
\end{eqnarray}
Thus non-accelerating perfect fluids are non-rotating or dust (i.e.
$p$ is constant). From this viewpoint, the class of non-rotating
dust forms the intersection of both possibilities, and in
\cite{Wylleman} it was shown that PM non-rotating dust spacetimes
(so-called anti-Newtonian universes~\cite{Maartens2}) do not exist.
Natural questions which then arise are (a) whether PM rotating dust
is allowed, and (b) whether non-accelerating and non-rotating PMpf's
exist. As said, we will mainly concentrate on question (b), which
will be dealt with in the next section. Regarding question (a) we
mention here that, just as in the non-rotating case
\cite{Maartens2}, PM rotating dust spacetimes are subject to at
least two chains of severe integrability conditions, built from the
consecutive propagations along $u^a$ of the constraint equations
(\ref{divE}) and (\ref{dotE}). A detailed analysis will be given
elsewhere, but it is clear that only a few distinct families of
solutions will be allowed, at most. In this section, we explicitly
show that the answer to (a) is negative for the subcases of Petrov
type D and of vanishing spatial 3-gradient $D_a\mu$ of the energy
density. The result for Petrov type D follows implicitly from
\cite{VdBWyll}, but the reasoning presented here provides a more
transparent and direct way to prove it. The result for $D_a\mu=0$
will be used for a further characterization of the new metric in the
next section.

In the proofs below, we will use that for $\dot{u}_a=0$ and $h_1=0$, the diagonal components of (\ref{dotsigma}) reduce to the single equation
\begin{eqnarray}
    \label{dotsD}\partial_0 h_2=\frac{1}{3}(h_2-2\t)h_2+\w_1^2.
\end{eqnarray}
Note that for Petrov type D, $\mu+p=0$ is not allowed \emph{an se} by the result of \cite{McIntosh1}.\\

{\bf Theorem 3.1} The class of PM type D dust spacetimes is empty.\\

{\bf Proof.} Suppose a PM type D dust spacetime $({\cal
M},g_{ab},u_a)$ exists. Taking an orthonormal eigenframe
${\cal B}$
of $H_{ab}$, we may assume that $H_{22}=H_{33}$. In this case
the (22), (33) and (23) components of (\ref{dotE}), (\ref{dotH}) immediately give
$q_1=-r_1,\s_{23}=0,n_{33}=n_{22}$ and $h_1=0$ (such that $h_3=-h_2$), while
the (12), and (13) components of (\ref{dotE}), (\ref{dotH}) together with the (2,3) components of (\ref{divH}) lead to
\begin{eqnarray}
& & W_2 = -\frac{1}{2}(\s_{13}+\w_2),\  W_3 = \frac{1}{2}(\s_{12}-\w_3) \\
& & 6H_{22}q_2+(\mu+p)(\s_{13}-3\w_2) = 0, \nonumber \\ \label{q2r3} & & 6H_{22}r_3+(\mu+p)(\s_{12}+3\w_3) = 0 .
\end{eqnarray}
Now from the (22), (33) and (23) components of (\ref{defH}) and from $\begin{pmatrix}  &3&\\0&1&3\end{pmatrix}+\begin{pmatrix}  &2&\\0&1&2\end{pmatrix}$ and
$\begin{pmatrix}  &2&\\0&3&1\end{pmatrix}-\begin{pmatrix}  &3&\\0&1&2\end{pmatrix}$ one gets
\begin{eqnarray}
    & & (\s_{12}+\w_3)q_2+(\s_{13}-\w_2)r_3=0,\nonumber \\
    & & (\s_{13}-\w_2)q_2-(\s_{12}+\w_3)r_3=0,
\end{eqnarray}
such that either $\s_{13}=\w_2$ and $\s_{12}=-\w_3$, or $q_2=r_3=0$; in the
latter case one has $\s_{13}=3\w_2,\s_{12}=-3\w_3$ by (\ref{q2r3}).
In both cases
the (22), (33) and (23) components of (\ref{dotsigma})
imply $(\w_3-\w_2)(\w_3+\w_2)=0$ and
$\w_2\w_3=0$, such that $\s_{12}=\s_{13}=\w_2=\w_3=0$. Under these
conditions, (\ref{divE}) yields
\begin{eqnarray}\label{divEtypeD}
    m_1=-18h\w_1,\quad m_2=m_3=0,
\end{eqnarray}
while from (\ref{dotw}) and (\ref{dottheta}) we obtain in the variables (\ref{thetaa}), (\ref{eqrefomega})
\begin{eqnarray}\label{dotw1sH}
    \partial_0\w_1&=&-2\t_{22}\w_1,\nonumber \\
     \partial_0 \t_{22} &=&\w_1^2-\frac{1}{6}(\mu+3p)-\t_{22}^2.
\end{eqnarray}
However, with (\ref{divEtypeD}) and $\w_1\neq 0\neq \mu+p$ one deduces from
$[\partial_2, \partial_3]\mu$ and (\ref{dotE}) that $\t_{22}\equiv \frac{1}{3}(h_2+\t)=0$. 
Taking two $\partial_0$-derivatives hereof and using (\ref{dotw1sH}), (\ref{dotmu}), $\dot{p}=0$ and $\mu+p\neq 0$, yields $\theta=h_2=0$ in contradiction with (\ref{dotsD}).$\hfill \square$.\\

{\bf Theorem 3.2} A non-accelerating PMpf with vanishing spatial 3-gradient of the energy density is non-rotating.\\

{\bf Proof.} Let $({\cal M},g_{ab},u_a)$ be a PMpf with
$\dot{u}_a=D_a\rho=0$. As non-accelerating perfect fluids are dust
or non-rotating, we may assume the Petrov type is I, since for
Petrov type D the result follows from Theorem 3.1. The  algebraic
vector constraint (\ref{divE}) and its covariant derivative along $u^a$
yield
\begin{eqnarray}
    \label{incompr} [\s,H]_a-3H_{ab}\w^b=0,\\
    \label{incompr_bis} 6\s_{c\langle a}{H^c}_{b\rangle }\w^b+\textrm{tr}(\s H)\w=
   \e_{abc}\w^b{H^c}_{d}\w^d.
\end{eqnarray}
Projected onto an orthonormal eigenframe ${\cal B}$ of $H_{ab}$, the
components of these equations form a system of 6 linear equations
in the variables $\s_{12},\s_{13},\s_{23},h_{2}$ and $h_{3}$,
parametrized by the $H_{\a\a}$ and $\w_a$, which can only be
consistent if the determinant of the so called extended system
matrix vanishes. Writing $x_\a=H_{\a+1\,\a+1}-H_{\a-1\,\a-1}$, this
yields
\begin{eqnarray*}\label{determinant}
x_1^4(x_2^2+x_3^2)\w_2^2\w_3^2 +\textrm{ cyclic terms } = 0.
\end{eqnarray*}
Hence, as the Petrov type is I, at least two of the spatial
components  $\w_\a$ must be zero, say $\w_2=\w_3=0$. Herewith
(\ref{incompr}), (\ref{incompr_bis}) imply $\s_{13}=\s_{12}=0$. Then $W_2=W_3=0$
follows from (\ref{dotH}) and (\ref{incompr_bis}) reduces to
$$\label{Incompr_bis} 6\s_{ab} H^{bc}\w_c-\s_{bc} H^{bc}\w_a=0.
$$
Taking a further covariant time derivative hereof, one finds
\begin{eqnarray}\label{dotdotdivE}
    & & (\s_{cd}\s^{cd}-3\w_c\w^c)H_{ab}\w^b-2\s_{cd}H^{cd}\s_{ab}\w^b \nonumber \\
    & & -\s_{bc}\s^{cd}{H^b}_{d}\w_a=0.
\end{eqnarray}
Suppose now $\w_1\neq 0$, such that ${\mathbf 0}\neq\w\propto {\mathbf e}_1$ is an eigenvectorfield of $H_{ab}$.
Eliminating $\s_{23}$ from the first components of (\ref{dotdotdivE}) and (\ref{incompr}) (the latter being given by $
\s_{23}x_1+\w_1(x_3-x_2)=0$), and then eliminating $h_{2}$, resp. $h_{3}$ from the result by means of (\ref{Incompr_bis}) (i.e. $h_2x_3+h_3x_2=0$), one finds
\begin{eqnarray*}
    & & \w_1^3x_2x_3(x_2-x_3)(4w_1^2x_3^2+h_3^2x_1^2)\\
    &=& \w_1^3x_2x_3(x_2-x_3)(4w_1^2x_2^2+h_2^2x_1^2)=0.
\end{eqnarray*}
Hence $x_3=x_2$ (i.e. $H_{11}=0$),
such that $h_1=\s_{23}=0$ by (\ref{Incompr_bis}) and  (\ref{incompr}), while the first component of
(\ref{dotH}) yields $W_1=-w_1/2$. Now the (11) and (1) components of (\ref{dotE}) and (\ref{divH})
respectively read
\begin{eqnarray}
    (n_{22}-n_{33})x_2-\frac{h_2}{3}(\mu+p)=0, \label{e:Eq4}\\
     (q_1+r_1)x_2+\w_1(\mu+p)=0,
\end{eqnarray}
which can only be consistent for $x_2\neq 0$ if
$\w_1(n_{22}-n_{33})+\frac{1}{3}(q_1+r_1)h_2=0$. On the other hand, the (23) component of (\ref{defH}) yields $\w_1(n_{22}-n_{33})-(q_1+r_1)h_2=0$, such that
$n_{22}=n_{33}$, whence $h_2=0$ from (\ref{e:Eq4}), again in contradiction
with (\ref{dotsD}).
$\hfill\square$ \\

{\bf Remark.} This generalizes
the result
of \cite{VdB1}.
The reasoning there was similar, but shortcuts were available due to
the non-existence of Petrov type D~\cite{McIntosh1} and type I($M^\infty$)~\cite{Brans} vacua.

 From \cite{Wylleman} and theorem 3.2 it follows:\\

{\bf Corollary.} PM dust spacetimes with vanishing spatial
3-gradient of the energy density do not exist.

\section{Algebraically general PMpf's}

Looking at (\ref{Ricciu}), (\ref{dotw}) and (\ref{Dp}), one sees
that \emph{non-rotating} (non-vacuum) perfect fluids in general obey
$\e_{abc}D^b\mu\,\dot{u}^c =\e_{abc}D^b\mu\, D^c p =0$, i.e., either
(a) the acceleration $\dot{u}_a$ and the spatial 3-gradient of the
pressure $D_a p$ are $\not = 0$ and proportional to the spatial 3-gradient of the
energy density $D_a\mu$, say $D_a p=fD_a\mu, f\neq 0$,
(b) $\dot{u}_a=D_a p=0$ or (c) $D_a\mu=0$. Hence, a non-rotating
perfect fluid exhibits a barotropic equation of state $p=p(\mu)$ if
and only if either
(1) the fluid has property (a) with $f=\frac{\partial p}{\partial \mu}=\frac{\dot{p}}{\dot{\mu}}\equiv
-\frac{\dot{p}}{(\mu+p)\t}\neq 0$, (2) the fluid has the properties (b) \emph{and} (c), or (3) it is dust (constant $p$, i.e.
$\frac{\partial p}{\partial \mu}=0$).

For non-rotating PMpf's, it was shown in \cite{Wylleman} that the
dust case (3) is impossible. On the other hand,  non-rotating
\emph{type D} PMpf's were fully classified in \cite{Lozanovski2}.
Such spacetimes were shown to belong to class (2) above and turned
out to be spatially homogeneous and locally rotationally symmetric
(LRS) of class III in the Stewart-Ellis classification.
The general metric form as well as the corresponding equation of
state were explicitly constructed, and the solutions satisfied the
energy conditions in an open region of spacetime. This result
motivates the investigation in the present section of the
existence of Petrov type I non-rotating PMpf's obeying (2). It is
proved that for non-rotating PMpf's (b) and (c) are actually
equivalent, and that non-rotating and algebraically general PMpf's
exist for which the spatial gradients of matter density and pressure
vanish. The solution is unique up to a constant rescaling of the
metric. This provides an affirmative answer to question (ii) at the
beginning of the previous section. The spacetime is characterized in
Theorem 4.2: it is found to be of Petrov type $I(M^{\infty}$),
has a degenerate shear tensor and is spatially homogeneous of
Bianchi type VI$_0$. Both the metric and equation of state are
constructed, and the physical behavior of the model is briefly
discussed. Finally, the solution is situated within a larger class
of perfect fluid models by dropping the purely magnetic condition.

\subsection{Characterization.}

Suppose (${\cal M},g_{ab},u_a$) is a Petrov type I non-rotating PMpf, and project the basic equations w.r.t.\ an orthonormal eigenframe ${\cal B}$ of  ${\s}_{ab}$, i.e., $\s_{\a+1\,\a-1}=0$. Then the $(\a\a)$ components of (\ref{defH}) become algebraic and very compact in the $n_\a$-variables:
\begin{eqnarray}\label{Haa}
    H_{\a\a}=n_{\a+1}h_{\a+1}-n_{\a-1}h_{\a-1}.
\end{eqnarray}
On the other hand, one can solve the $(0\a)$ field equations and the $(\a+1\,\a-1)$ components of (\ref{defH}) for $\partial_\a h_{\a+1},\partial_\a h_{\a-1}$, giving
\begin{eqnarray}
    \label{eah1}\partial_\a h_{\a+1}&=&-\partial_\a\t-H_{\a+1\a-1}-r_\a h_{\a-1}-2q_\a h_{\a+1},\\
    \label{eah2}\partial_\a h_{\a-1}&=&\partial_\a\t-H_{\a+1\a-1}+2r_\a h_{\a-1}+q_\a h_{\a+1},
\end{eqnarray}
while (\ref{dotsigma}) and (\ref{dottheta}) combine to prescriptions for the evolution of the variables (\ref{thetaa}):
\begin{eqnarray}
    \partial_0 \t_{\a\a}&=&-\t_{\a\a}^2 +\partial_\a\dot{u}_a+\dot{u}_a^2+q_{\a+1}\dot{u}_{\a+1}-r_{\a-1}\dot{u}_{\a-1}\nonumber \\
    \label{dotexpaa} & & -\frac{1}{6}(\mu+3p).
\end{eqnarray}
As shown below and in the appendix, we may take $W_\a=0$ for the cases $D_a\rho=0$ and $D_a p=\dot{u}_a=0$ under study. Then the $(\a\a)$ components of (\ref{dotH}) reduce to
\begin{eqnarray}\label{dotHaa}
    \partial_0 H_{\a\a} &=&-\t H_{\a\a}+\frac{1}{2}(h_{\a-1}-h_{\a+1})H_{\a\a}\nonumber \\
    & & +\frac{1}{2}h_\a(H_{\a-1\,\a-1}-H_{\a+1\,\a+1}).
\end{eqnarray}
Inserting (\ref{eah1})-(\ref{eah2}) into $\begin{pmatrix}&\a-1&\\0&\a&\a-1\end{pmatrix}$ and $\begin{pmatrix}&\a+1&\\0&\a&\a+1\end{pmatrix}$ one gets
\begin{eqnarray}\label{e0qr}
    \partial_0q_\a&=&
    \frac{1}{3}(h_\a-h_{\a+1}-\t)(q_\a-u_\a)-H_{\a+1\,\a-1},\nonumber
    \\
    \partial_0r_\a&=&
    \frac{1}{3}(h_{\a-1}-h_{\a}-\t)(r_\a+u_\a)-H_{\a+1\,\a-1}
\end{eqnarray}
while $\begin{pmatrix}&\a&\\0&\a+1&\a-1\end{pmatrix}$
yields
\begin{eqnarray}
    \label{e0nn}\partial_0 n_{\a\a}=-\frac{1}{3}\t n_{\a\a}+\frac{2}{3}(h_{\a-1}-h_{\a+1})n_{\a\a},
\end{eqnarray}
or translated for $n_\a$ instead of $n_{\a\a}$:
\begin{eqnarray}\label{e0n}
\partial_0n_\a&=&-\frac{1}{3}\t
n_\a-\frac{1}{3}(h_{\a-1}-h_{\a+1})n_\a \nonumber \\
& &
+h_{\a+1}n_{\a+1}-h_{\a-1}n_{\a-1}.
\end{eqnarray}

We will make use of the following\\

{\bf Lemma 4.1} W.r.t.~a shear-eigenframe of a non-rotating PMpf:\\
(a) at most one of the $n_\a$ vanishes;\\
(b) at most one of the $h_{\a}$ vanishes;\\
(c) if $\dot{u}_a=0$, then $a h_2 +b h_3=0$ for constant $a,b$ ($(a,b)\neq(0,0)$) implies $h_1h_2h_3=0$, i.e.~the shear tensor is degenerate.\\

{\bf Proof.} (a) Immediately follows from (\ref{e0n}) and (\ref{Haa}). \\
(b) This is the statement that for non-rotating PMpf's the shear
tensor cannot vanish~\cite{Trumper}, as is immediately seen from
(\ref{defH}) in covariant form (or from (\ref{Haa}) and
(\ref{eah1})-(\ref{eah2}) in tetrad form). \\
(c) For $\dot{u}_\a=0$,
the difference of the $(\a+1\, \a+1)$ and $(\a-1\, \a-1)$ components of (\ref{dotsigma}) reads
\begin{eqnarray}
    \label{e0hu}\partial_0h_\a=
    -\frac{1}{3}(h_{\a+1}-h_{\a-1}+2\t)h_\a.
\end{eqnarray}
 From this and $h_1+h_2+h_3=0$, one deduces $\partial_0(ah_2+bh_3)+\frac{1}{3}(2\t+h_1-h_2)(ah_2+bh_3)=ah_1h_2$, from which the result follows.
$\hfill\square$\\

{\bf Theorem 4.2} For algebraically general PMpf's, 
any two of the following three conditions imply the third:
\begin{enumerate}
    \item[(i)] the fluid congruence is non-rotating; 
    \item[(ii)] the fluid is non-accelerating (i.e.~the spatial gradient of the pressure vanishes); 
    \item[(iii)] the fluid has a vanishing spatial gradient of the matter density. 
\end{enumerate}
{\bf Proof.}\ \\
$(i),(ii)\Rightarrow (iii)$. See appendix.\\
$(ii),(iii)\Rightarrow (i)$. This was the content of Theorem 3.2.\\
$(iii),(i)\Rightarrow (ii)$. When $\partial_\a\mu=0$ one deduces
from $[\partial_0, \partial_\a]\mu$ (or (\ref{comTD})) and (\ref{Dp}) that also
$\partial_\a\t=0$. With $H_{\b+1\,\b-1}=0$, (\ref{dotE}) becomes
algebraic:
\begin{eqnarray}\label{kwak}
    & & (H_{\a\a}-H_{\a-1\,\a-1})n_{\a+1\,\a+1}+(H_{\a\a}-H_{\a+1\,\a+1})n_{\a-1\,\a-1}\nonumber \\
    & & -\frac{\mu +p}{6}(h_{\a+1}-h_{\a-1})=0
\end{eqnarray}
and consists of two independent equations. Elimination of $\mu +p$
yields
\begin{eqnarray}\label{krak}
    (h_{2}-h_{3})(H_{22}-H_{33})n_1+ \textrm{ cyclic terms } = 0
\end{eqnarray}
 From (\ref{Haa}) one deduces
\begin{eqnarray}
\label{naHa} & & n_\a H_{\a\a} = n_\a(n_{\a+1}h_{\a+1}-n_{\a-1}h_{\a-1}),\\
& & (n_1n_2+n_2n_3+n_3n_1)h_{\a}=\nonumber
\\
\label{nana} & & n_{\a-1}H_{\a-1\,\a-1}-n_{\a+1}H_{\a+1\,\a+1}.
\end{eqnarray}
If $X\equiv n_1n_2+n_2n_3+n_3n_1$ vanishes, then $Y\equiv
n_1H_{11}=n_2H_{22}=n_3H_{33}$ from (\ref{nana}) and $\partial_0
X=-2Y=0$ from (\ref{naHa}), which is contradictory to lemma 4.1 (a).
Hence $X\neq 0$ and one can use (\ref{nana}) to eliminate the
$h_\a$. Doing this for (\ref{krak}) one obtains $(n_1+n_2+n_3)F=0$,
with
\begin{eqnarray*} F\equiv n_1H_{11}(H_{22}-H_{33})+ \textrm{ cyclic
terms }.
\end{eqnarray*}
This leaves two cases
\begin{enumerate}
    \item[(I)] $F=0$:\\
     Calculating $\partial_0 F+\frac{7}{3}\t F$ and substituting for $h_\a$ one gets another polynomial equation $G=0$ in
    $n_\a$ and $H_{\a\a}$. Substituting $H_{\a+1\,\a+1}=-H_{\a\a}-H_{\a-1\,\a-1}$ and taking the resultant of $F$ and $G$ w.r.t. $H_{\a-1\,\a-1}$
    yields
\begin{eqnarray*}
&& H_{\a\a}(n_1-n_2)(n_2-n_3)(n_3-n_1)\\
&& \times (n_1n_2+n_2n_3+n_3n_1)^2=0.
\end{eqnarray*}
 Hence e.g.~$n_3=n_2$. Inserting this in $F=0$ yields $H_{11}(H_{22}-H_{33})(n_1-n_2)=0$. If $H_{11}$ were
 zero then $G=0$ would read $H_{22}^3n_2^2=0$, hence $n_2=n_3=0$, again in contradiction with lemma 4.1 (a); if $n_1-n_2$
 were zero then $G=0$ would read $n_2^2(H_{11}-H_{22})(H_{22}-H_{33})(H_{33}-H_{11})=0$. Thus $F=0$ leads to Petrov type D (and hence to the models found in \cite{Lozanovski2}),
contrary to our type I assumption.
    \item[(II)] $n_1+n_2+n_3=0$:\\
 Applying $\partial_0$ to this equation one finds
\begin{eqnarray}\label{naha}
    (h_2-h_3)n_1+(h_3-h_1)n_2+(h_1-h_2)n_3=0
\end{eqnarray}
or, by (\ref{nana}),
\begin{eqnarray}\label{naMa}
    P_1\equiv n_1^2H_{11}+n_2^2H_{22}+n_3^2H_{33}=0.
\end{eqnarray}
Calculating $\partial_0 P_1+\frac{5}{3}\t P_1$ and again eliminating the
$h_\a$ by (\ref{nana}) one obtains a further polynomial equation
$P_2$ in the $n_{\a}$ and $H_{\a\a}$. Substituting
$H_{\a+1\,\a+1}=-H_{\a\a}-H_{\a-1\,\a-1}$ and eliminating $H_{\a-1\,\a-1}$ from $P_1$ and $P_2$ now yields
\begin{equation}
H_{\a\a}^2n_1n_2n_3(n_1n_2+n_2n_3+n_3n_1)^2=0.
\end{equation}
Hence e.g.~$n_1=0$,
such that $n_2+n_3=n_{11}=0$, $h_1=0$ from (\ref{naha}) and
$H_{11}=0$ from (\ref{naMa}).
 From $h_1=0$ and (\ref{eah1}), (\ref{eah2}) one deduces $q_1=-r_1$
and $q_2=0,r_3=0$. Now $\partial_0 q_2=\partial_0 r_3=0$ yields
$\dot{u}_2(2h_2-\t)=\dot{u}_3(2h_2-\t)=0$ by means of (\ref{e0qr}). \\
Suppose that $\dot{u}_2$ and $\dot{u}_3$ are non-zero, such that
$2h_2-\t\equiv -3\t_{11}=0$.
Then 
$\partial_1(2h_2-\t)=0$ yields $h_2(r_1-2q_1)=0$ by (\ref{eah1}) and
$\partial_1\t=0$, such that $q_1=r_1=0$ by lemma 4.1 (b). Now
$\partial_0 r_1 =0$ yields $h_2\dot{u}_1=0$ by (\ref{e0qr}) and hence,
again by lemma 4.1 (b), $\dot{u}_1=0$. Then however, by (\ref{dotexpaa}) for
$\a=1$, one would have $\mu+3p=0$, leading to $\partial_\a
p=\dot{u}_\a=0$, contradictory to the assumption $(\dot{u}_2,
\dot{u}_3)\not = (0,0)$. Hence $\dot{u}_2=\dot{u}_3=0$. Finally, the (23) component of (\ref{dotsigma})
becomes $\dot{u}_1(n_2-n_3)=0$, such that also $\dot{u}_1=0$ by lemma 4.1 (b), and
the fluid is non-accelerating.
\end{enumerate}

\subsection{Existence and properties.}\label{Exprop}

 From the proof of theorem 4.2 it follows that for non-rotating,
algebraically general PMpf's with vanishing spatial gradients of
matter density and pressure, one of the eigenvalues of $H_{ab}$ is
identically zero, such that the Petrov type is I($M^\infty$) in the
extended Arianrhod-McIntosh classification. The
$H_{ab}$-eigenframe
$\{{\mathbf u},{\mathbf e}_\a\}$ being also an eigenframe of the
shear tensor, the latter is degenerate in the plane perpendicular to
the 0-eigendirection of $H_{ab}$. When ${\mathbf e}_1$ spans this
eigendirection, i.e.~for $H_{11}=h_1=0$ (such that
$H_{33}=-H_{22}\neq 0,h_3=-h_2\neq 0$), the following equations have
by now been established:
\begin{eqnarray}
    \label{tensorial1}\partial_\a p &=& \partial_\a\mu=\partial_\a\t=0,\\ 
    \label{tensorial2}\dot{u}_\a &=& \w_\a=\s_{\a+1\,\a-1}=0,\\
    \label{Wa}W_\a &=& 0,\\
    \label{n} n_1 &=& n_2+n_3=0 \quad(\textrm{i.e. } {n^\a}_\a=n_{11}=0),\\
    \label{qr}q_1+r_1 &=& q_2=r_3=0,
\end{eqnarray}
while (\ref{Haa}) and (\ref{kwak}) reduce to
\begin{eqnarray}
\label{tak} H_{22} &=& n_2h_2,\\
\label{snak}    n_2^2 &=& \frac{1}{6}(\mu+p).
\end{eqnarray}
Now from (\ref{qr}) and (\ref{eah2}), (\ref{eah1}) one sees that
$\partial_2 h_2=\partial_3 h_2=0$, while with (\ref{tensorial1}),
(\ref{tensorial2}) and (\ref{dottheta}) one immediately deduces from
$[\partial_0,\partial_1]\t$ that also $\partial_1 h_2 =0$, whence $q_1=r_1=0$ by
(\ref{qr}) and (\ref{eah1}). From the $(\a+1\,\a-1)$ field equations and $\begin{pmatrix} &2&\\1&2&3 \end{pmatrix}$, $[\partial_1, \partial_2]n_2$ and $[\partial_3,\partial_1]n_2$ one deduces
$q_3=r_2=0$ and $\partial_\a n_2=0$
(alternatively, $\partial_\a n_2=\partial_\a H_{22}=0$ follows from
(\ref{tak}) and (\ref{snak}), whence $r_2=q_3=0$
by (\ref{divH}), but the previous reasoning 
is more generally valid, cf.~infra). Finally, the
$(\a\a)$ field equations and (\ref{snak}) are equivalent to
\begin{eqnarray}
   \label{n2kwad}n_2^2&=&\frac{1}{2}\t_{22}(\t_{11}-\t_{22}),\\
     \label{mu}\mu&=&\frac{3}{2}\t_{22}(\t_{11}+\t_{22}),\\
     \label{p}p&=&\frac{3}{2}\t_{22}(\t_{11}-3\t_{22}),
\end{eqnarray}
where, in view of the construction of the metric below, we
re-introduced the variables $\t_{11}$ and $\t_{22}=\t_{33}$. By
virtue of (\ref{dotHaa}), (\ref{e0n}), (\ref{dotexpaa}) and
(\ref{dotmu}), the relations (\ref{tak}), (\ref{n2kwad}) and
(\ref{mu}) are consistent under propagation along the matter flow
lines, whereas applying $\partial_0$ to (\ref{p}) yields an
expression for $\dot{p}$ in $\t_{11}$ and $\t_{22}$:
\begin{eqnarray}
    \dot{p}=-3\t_{22}(\t_{11}-\t_{22})(\t_{11}-2\t_{22}).
\end{eqnarray}
With the above specifications, all basic equations are satisfied and
consistent, implying that corresponding
solutions exist. Since all
$\partial_\a$-derivatives of the commutator coefficients vanish the
spacetimes will be spatially homogeneous, while $q_\a=r_\a=0$ and
(\ref{n}) imply that the Bianchi type is VI$_0$ (see
\cite{EllisMacCallum}).

\subsection{Metric, uniqueness and equation of state}

With (\ref{tensorial2})-(\ref{n}), $q_\a=r_\a=0$ and $\t_{22}=\t_{33}$, it can be easily checked that the normalized vector fields
\begin{eqnarray}
    && {\mathbf e}_0\equiv{\mathbf u},\quad {\mathbf e}_1,\nonumber \\
    && {\mathbf e}_{2^\prime}=\frac{1}{\sqrt{2}}({\mathbf e}_2+{\mathbf e}_3),
    \quad {\mathbf e}_{3^\prime}=\frac{1}{\sqrt{2}}({\mathbf e}_2-{\mathbf e}_3)
\end{eqnarray}
are hypersurface-orthogonal and by (\ref{commrel}) obey
\begin{eqnarray}
    && [{\mathbf e}_0,{\mathbf e}_1]=-\t_{11}{\mathbf e}_1,\nonumber \\
   &&  \label{com0}[{\mathbf e}_0,{\mathbf e}_{2^\prime}]=
    -\t_{22}{\mathbf e}_{2^\prime}, \quad [{\mathbf e}_0,{\mathbf e}_{3^\prime}]=-\t_{22}{\mathbf e}_{3^\prime}, \\
   && [{\mathbf e}_1,{\mathbf e}_{2^\prime}]=n_2{\mathbf e}_{2^\prime},
    \quad [{\mathbf e}_1,{\mathbf e}_{3^\prime}]=-n_2{\mathbf e}_{3^\prime},\nonumber \\
    && \label{comspat} [{\mathbf e}_{2^\prime},{\mathbf e}_{3^\prime}]=0.
\end{eqnarray}
Following $\S 6$ of \cite{EllisMacCallum} and noting that $n_2\neq 0$ because of (\ref{tak}), we may choose coordinates $t^\prime,x^\prime,y^\prime,z^\prime$ such that
\begin{eqnarray}
       && {\mathbf e}_0=\frac{\partial}{\partial t^\prime}, \quad {\mathbf e}_1=n_2(t^\prime)\frac{\partial}{\partial x^\prime},\nonumber \\
       && {\mathbf e}_2^\prime=B(t^\prime)e^{x^\prime}\frac{\partial}{\partial y^\prime}, \nonumber \\
       && {\mathbf e}_3^\prime=B(t^\prime)e^{-x^\prime}\frac{\partial}{\partial z^\prime}.
\end{eqnarray}
Relations (\ref{com0}) and (\ref{comspat}) are satisfied if and only
if
\begin{eqnarray}\label{dynsys}
    \frac{d \ln{n_2}}{dt^\prime}=-\t_{11},\quad \frac{d \ln{B}}{dt^\prime}=-\t_{22}.
\end{eqnarray}
By (\ref{mu}), (\ref{p}) and (\ref{dotexpaa}), the autonomous
dynamical system for the fundamental variables $\t_{11}$ and
$\t_{22}$ becomes
\begin{eqnarray}
\label{kraak1}  \frac{d\t_{11}}{dt^\prime}&=&(\t_{22}-\t_{11})(\t_{11}+2\t_{22}),\\
\label{kraak2}
\frac{d\t_{22}}{dt^\prime}&=&(\t_{22}-\t_{11})\t_{22}.
\end{eqnarray}
As $\t_{11}=\t_{22}$ and $\t_{22}=0$ are not allowed by Lemma 1 (b)
and e.g.~(\ref{mu})+(\ref{p}), one derives from (\ref{kraak1}) and
(\ref{kraak2}) that
\begin{eqnarray}\label{kraak3}
    \t_{11}&=&(2\ln{\t_{22}}-C_0+1)\t_{22}\quad\textrm{or}\nonumber \\
     u&\equiv& \frac{\t_{11}}{\t_{22}}-1=2\ln{\t_{22}}-C_0,
\end{eqnarray}
with $C_0$ an integration constant. Using $u$ as time variable, this
yields
\begin{eqnarray}\label{t1122}
    \t_{22}=\frac{C}{2}\exp{(\frac{u}{2})},\quad \t_{11}=\frac{C}{2}\exp{(\frac{u}{2})}(u+1),
\end{eqnarray}
with
\begin{eqnarray*}
C\equiv 2\exp{(\frac{C_0}{2})}.
\end{eqnarray*}
 From (\ref{kraak2}) and (\ref{kraak3}) one obtains 
$du=-2\t_{22}udt^\prime=-Cu\exp{(\frac{u}{2})}dt^\prime$, such that (\ref{dynsys}) may be integrated to give
\begin{eqnarray}\label{n2B}
    n_2(u)=C_1\,e^{\frac{u}{2}}u^{\frac{1}{2}},\quad B(u)=C_2u^{\frac{1}{2}},
\end{eqnarray}
with $C_1$ and $C_2$ non-zero constants. From (\ref{n2kwad}) one
finds $8C_1^2=C^2$, and after a coordinate change
$u=2e^{-t},x^\prime=x/2,y^\prime=\sqrt{2}\,C_2y,
z^\prime=\sqrt{2}\,C_2z$ the metric reads
\begin{eqnarray}\label{metric}
    C^2ds^2&=&\exp(-2e^{-t})(-dt^2+e^t dx^2)\nonumber \\
    & & +e^t(e^{-x}dy^2+e^{x}dz^2).
\end{eqnarray}
Thus we find a unique spacetime up to a constant rescaling.
The only non-zero components of the projected Weyl and shear tensors
are
\begin{eqnarray}
    H_{2^\prime3^\prime}(t)&=&H_{22}(t)\nonumber \\
    &=& -\frac{C^2}{2}\exp(-\frac{3}{2}t+2 e^{-t}),\\
    \s_{2^\prime2^\prime}(t)&=&\s_{3^\prime3^\prime}(t)=-\frac{1}{2}\s_{11}(t)\nonumber \\
    &=&-\frac{1}{3}C\exp(-t+e^{-t}),
\end{eqnarray}
while the scalar expansion rate, energy density
and pressure are given by
\begin{eqnarray}
    \t(t)&=& C\exp{(e^{-t})}(e^{-t}+\frac{3}{2}),\\
    \label{mut} \mu(t) &=& \frac{3C^2}{4}\exp(2 e^{-t})(e^{-t}+1),\\
    \label{pt} p(t) &=& \frac{3C^2}{4}\exp(2 e^{-t})(e^{-t}-1).
\end{eqnarray}
Note from (\ref{mut}) and (\ref{pt}) that constant $p$ is not
allowed, which could be deduced more directly from (\ref{p}) and
(\ref{kraak1})-(\ref{kraak2}); together with theorem 4.2 this
provides an alternative proof for the non-existence of irrotational
PM dust~\cite{Wylleman}. One may eliminate $t$ from (\ref{mut}) and
(\ref{pt}) via $\mu-p=\frac{3C^2}{2}\exp{(2e^{-t})}>0$, which yields
the following equation of state
\begin{eqnarray}\label{eqstate}
    \mu+p=\frac{1}{2}(\mu-p)\ln{\left(\frac{2(\mu-p)}{3C^2}\right)}.
\end{eqnarray}
We have $u^a=C\exp(e^{-t})\frac{\partial}{\partial t}$, which is
future directed for $C>0$. We conclude that (\ref{metric}) is the
metric of a perfect fluid model, which starts off with a stiff
matter-like big-bang singularity at a finite proper time in the past
(corresponding to $t=-\infty$, with $\lim_{t\rightarrow -\infty}\mu
/ p =1$) and which expands indefinitely towards
an Einstein space with
$\mu(\infty)=\frac{3C^2}{4},p(\infty)=-\frac{3C^2}{4},\t(\infty)=\frac{3C}{2}$.
Note that the dominant energy condition
$\mu>0,-\mu<p<\mu$ is satisfied throughout spacetime,
whereas at $t=0$ (i.e.~after a proper time
$\int_1^{\infty}\exp(-u)\frac{du}{u}$) $p$ becomes negative.

We conclude with:\\

{\bf Theorem 4.3.} Up to a constant rescaling of the metric, there
exists a unique purely magnetic perfect fluid which satisfies any
two of the three properties of Theorem 4.2. This fluid is
non-rotating and has vanishing spatial gradients of energy density and pressure, with the line element given by (\ref{metric}) and the equation of state by (\ref{eqstate}). It is
orthogonally spatially homogeneous of Bianchi type VI$_0$. The Petrov type
is I($M^\infty$) in the extended Arianrhod-McIntosh classification, the
shear tensor being degenerate in the plane perpendicular to the
0-eigendirection of the projected Weyl tensor.

\subsection{Relaxing the purely magnetic condition.} We want to indicate here how the algebraically
general purely magnetic spacetime of theorem 4.2 (and at the same
Lozanovski's type D class~\cite{Lozanovski2}) naturally fits into a
wider class of perfect fluid models. More precisely, we drop the
purely magnetic condition and look at the class ${\cal A}$ of
non-vacuum, non-conformally flat, non-rotating perfect fluids $({\cal M},g_{ab},u^a)$ which have a degenerate shear tensor $\s_{ab}\neq 0$ and vanishing spatial gradients of energy density and pressure.
As $\w_a=\dot{u}_a=D_a p=D_a\mu=0$, one derives $D_a\t=0$ from (\ref{dotmu}) and
(\ref{comTD}) with $f=\mu$,  after which $D_a(\s_{bc}\s^{bc})=0$
follows from (\ref{dottheta}) and (\ref{comTD}) with $f=\theta$. The shear being
degenerate, $\s_{bc}\s^{bc}$ is the only independent scalar which
may be built from $\s_{ab}$. As $\dot{u}_a=\w_a=0$, we may choose a
time coordinate $t$ such that ${\mathbf u}=\frac{\partial}{\partial t}$:
$\mu,p,\t$ and $\s_{ab}\s^{ab}$ depend then on $t$ only and it
follows that ${\cal A}$ is part of the class of so called
\emph{kinematically homogeneous} perfect fluids, defined and studied
in \cite{VdB3}. It follows from the analysis there that for any
member of ${\cal A}$, an eigenframe ${\cal
B}=\{{\mathbf u},{\mathbf e}_\a\}$ of $\s_{ab}$ exists, for which
(\ref{tensorial1})-(\ref{Wa}) and (\ref{qr}) hold, the shear being
degenerate in the $({\mathbf e}_2,{\mathbf e}_3)$ plane
($h_1=h_2+h_3=0$).
Herewith, it follows from (\ref{dotsigma}) and (\ref{defH}) that $\s_{ab},E_{ab}$ and $H_{ab}$ diagonalize
simultaneously in ${\cal B}$. Thus the fluid congruence $u^a$ is Weyl principal and, as follows from the introduction, purely magnetic (purely electric) spacetimes of ${\cal A}$ automatically have a PM (PE) Weyl tensor \emph{w.r.t.\ $u^a$}. From the (22) and (33) components of (\ref{dotsigma}) and (\ref{defH}) one obtains
\begin{equation}\label{EH22}
    E_{22}=E_{33},
    \quad H_{22}=-h_2n_3,\quad H_{33}=-h_2n_2.
\end{equation}
Also, the difference of the (22) and (33) components of the field equations reduces to
$n_1(n_2-n_3)=0$.

For $n_2=n_3$ the Petrov type is D, and it
was further deduced in \cite{VdB3} that the corresponding models are
spatially homogeneous and LRS III. Note from (\ref{EH22}) that any PE model in ${\cal A}$ is of Petrov type D (with
$n_2=n_3=0$) and was proved \cite{Barnes2} to belong to the Szekeres-Szafron family \cite{Szekeres, Szafron1,
Szafron2,Barrow}. This family is here characterized as the PE subclass of ${\cal A}$, whereas Lozanovski's family \cite{Lozanovski2} forms precisely the PM type D subclass of ${\cal A}$.

For $n_1=0$ on the other hand, one reobtains (\ref{n}) from \[\begin{pmatrix} &3&\\0&1&2\end{pmatrix} +\begin{pmatrix} &2&\\0&3&1\end{pmatrix}-\begin{pmatrix} &1&\\0&2&3\end{pmatrix},\]
while
$H_{11}=0,H_{33}=-H_{22}\neq 0$ then follows from (\ref{EH22}). For $n_2=n_3=0$ one obtains a subclass of the PE type D models of ${\cal A}$, while the Petrov type is I if and only if $n_2\neq 0$. Moreover, the essentially unique spacetime of theorem 4.3 can now be characterized alternatively as the unique type I$(M^\infty)$ member of ${\cal A}$; in this respect, also note that the invariants (\ref{eq:J}) and (\ref{eq:M}) become
\begin{eqnarray*}
    J=-2E_{22}(E_{22}^2+H_{22}^2),\quad M=-\frac{2H_{22}^2(9E_{22}^2+H_{22}^2)^2}{9E_{22}^2(E_{22}^2+H_{22}^2)}.
\end{eqnarray*}
Further, one derives that $q_\a=r_\a=\partial_\a n_2=0$ in this case, just as in section \ref{Exprop} (following \cite{VdB3}).
Thus the corresponding
spacetimes are spatially homogeneous ($\partial_\a\equiv0$) of
Bianchi type VI$_0$. The surviving equations describe the so-called
``degenerate shear'' subclass $S_1^+(VI_0)$ of the Bianchi $VI_0$
family\cite{WainwrightEllis},
\begin{eqnarray}
    \label{E22}E_{22} &=& \frac{2}{3}n_2^2+\frac{1}{9}h_2(h_2+\theta),\\
    \label{H22}H_{22} &=& h_2n_2,\\
    \label{muu}\mu &=& -n_2^2+\frac{\t^2-h_2^2}{3},\\
    \label{e0HH}\frac{d\t}{dt} &=& -\frac{1}{2}(\t^2+h_2^2-n_2^2+3p),\\
    \label{e0h2}\frac{dh_2}{dt} &=& -\t h_2-2n_2^2,\\
    \label{e0n2}\frac{dn_2}{dt} &=& -\frac{1}{3}(\t-2h_2)n_2.
\end{eqnarray}
The algebraic relations (\ref{E22})-(\ref{muu}) are consistent under
evolution by (\ref{e0HH})-(\ref{e0n2}) and the Bianchi propagation
equations. Hence for every choice of the free function $p(t)$, there
is a family of solutions to the Einstein equations corresponding
with the dynamical system (\ref{e0HH})-(\ref{e0n2}). When a specific
barotropic equation of state $p=p(\mu)$ is assumed, this system
becomes autonomous by (\ref{muu}); the case $p=-\Lambda$ is of
particular interest since a family of irrotational dust spacetimes is
then obtained, but an explicit integration is not known. Another
possibility to extract a subfamily is to impose conditions on the
Weyl tensor, e.g.~$E_{22}=bH_{22}$ for constant $b$. By (\ref{E22})
and (\ref{H22}) this equation turns (\ref{e0h2})-(\ref{e0n2}) into
an autonomous dynamical system in $h_2$ and $n_2$, parametrized by
$b$, the exact solution of which can be given in terms of elliptic
integrals. Alternatively, $E_{22}=bH_{22}$ and its time evolution
ensure that $n_2$ and $p$ become algebraically dependent of $h_2$
and $\t$, which turns (\ref{e0HH})-(\ref{e0h2}) into an autonomous
system. The case $b=0$ eventually yields the purely magnetic
spacetimes (\ref{metric}).
Finally note that imposing \emph{both} a specific equation of state and a relation of the form $E_{22}=bH_{22}$,
would generically force the mentioned dynamical systems to be inconsistent,
which clarifies again the result for PM irrotational dust \cite{Wylleman}.

\section{Conclusion}

Perfect fluid solutions of the Einstein field equations were
considered, for which the Weyl tensor is purely magnetic with
respect to the fluid velocity $u^a$. We first generalized the
results~\cite{Wylleman} on the non-existence of the so-called
anti-Newtonian non-rotating dust universes, to the case of rotating
dust which is either of Petrov type D, or which has a vanishing
spatial gradient of the matter density. These results, in
combination with some ongoing work on rotating dust (for which the further subcase of degenerate shear has by now been dealt with) lead us to conjecture that purely
magnetic dust spacetimes do not exist.\\
Motivated by the existence of non-rotating purely magnetic perfect
fluids of Petrov type D, for which both the spatial gradients of
energy density and pressure vanish~\cite{Lozanovski2}, we studied
the algebraically general case and demonstrated that any two of the
following conditions implies the third: (i) the fluid congruence is
non-rotating, (ii) the fluid is non-accelerating, (iii) the fluid
has a vanishing spatial gradient of the matter density. For purely
magnetic perfect fluids satisfying these conditions it turns out
that the magnetic part of the Weyl tensor has one 0 eigenvalue and
that the shear tensor is degenerate in the plane orthogonal to the
corresponding eigenvector.  The unique (up to constant rescaling)
solution satisfying these conditions is an orthogonally spatially
homogeneous perfect fluid of Bianchi type $VI_0$ which has a big
bang singularity, starts of as stiff fluid and asymptotically
evolves towards an Einstein space.

Finally notice that purely magnetic spacetimes of Petrov type D or $I(M^\infty)$, as considered in the present work, cannot be conformally mapped to vacuum spacetimes, by the results of \cite{McIntosh1} and \cite{Brans} and the fact that the Weyl tensor is preserved under such mappings. The question remains open whether type $I(M+)$ purely magnetic spacetimes exist at all.\\

An overview of the results obtained in this paper for Petrov type I is given in table I.\\
\begin{table*}
\Tree [.\fbox{type I PMpf's}
[.{$\dot{u}\not =0$} {$\omega \not=0$ \\ ?} [.{$\omega =0$} {$D\mu \neq 0$\\ ?} {$D\mu =0$\\ $\nexists$} ] ]
[.{$\dot{u} =0$} [.{$\omega =0$ \\ $\Downarrow$ \\ $D\mu =0$} {$p=const$\\ A \\ (\ref{metric})} {$p\neq const$\\ $\nexists$} ] [.{$\omega \not=0$ \\ $\Downarrow$ \\ $p=const$} {$D\mu \neq 0$\\ conjecture:\\ $\nexists$} {$D\mu =0$\\ $\nexists$} ] ]
]
\caption{Results obtained for PMpf's of Petrov type I: $\nexists$ and A indicate respectively that no solutions exist or that all solutions are known.}
\end{table*}

\textbf{Note added in proof}: it has been brought to our attention that the same metric of section 4.3 has been independently found by C. Lozanovski.

\begin{acknowledgements}
Lode Wylleman is a research assistant supported by
the Fund for Scientific Research Flanders(F.W.O.).
\end{acknowledgements}

\appendix*
\section{}

We provide here the quite long and technical proof for the part $(i),(ii)\Rightarrow (iii)$ of proposition 4.2, stating that for non-rotating and non-accelerating type I PMpf's the spatial 3-gradient of the energy density vanishes. Projection w.r.t.\ an orthonormal eigenframe ${\cal B}$ of $\s_{ab}$ turns out to be favorable, and we actually prove the result regardless of the Petrov type. We will split the proof in two cases: degenerate and non-degenerate shear.
For conciseness we will write $\rho\equiv\mu+p$ and we add $m_\a\equiv\partial_\a\mu=\partial_\a\rho,z_\a\equiv\partial_\a\t$ as variables. Note that we may exclude $\rho=0$ by the result of \cite{VdB1} or \cite{VdB2}.

With $\s_{\a+1\,\a-1}=0$, the $(\a+1\,\a-1)$ component of (\ref{dotsigma}) reads $W_\a h_\a=0$, such that $W_\a=0$ when the shear tensor is non-degenerate;
for degenerate shear, say $h_1=0$, only $W_2=W_3=0$ follows, but we may then partially fix the frame by taking $W_1=0$ (hereby leaving the freedom of performing rotations about an angle $x$ in the $({\mathbf e}_2,{\mathbf e}_3)$-plane satisfying $\partial_0 x=0$), such that also here the triad $\{{\mathbf e}_\a\}$ is Fermi-propagated along the fluid flow \cite{MacCallum}. Hence the equations (\ref{Haa})-(\ref{e0hu}) are valid.
Further, the Bianchi constraint divE$_\a$ translates to
\begin{equation} \label{Qa}
Q_\a\equiv  m_\a+3h_\a H_{\a+1\,\a-1}=0.
\end{equation}
On applying $\partial_0$ to $Q_\a$ and eliminating $H_{\a+1\,\a-1}$ by means of $Q_\a$,
one gets
\begin{equation}\label{e0m}
    \partial_0 m_\a=-\frac{5}{3}\t m_\a+\frac{1}{6}(h_{\a+1}-h_{\a-1})m_\a.
\end{equation}
Herewith, $[\partial_0 , \partial_{\a}]\rho$
becomes
\begin{equation}\label{Ra}
    R_\a \equiv\,\rho z_\a-\frac{1}{6}(2\t+h_{\a+1}-h_{\a-1})m_\a = 0
\end{equation}
while
$[\partial_0 , \partial_{\a}]\theta$
results in an expression for $\partial_0 z_\a$:
\begin{eqnarray}\label{e0z}
    \partial_0 z_\a &=& -\t z_\a +(h_{\a+1}-h_{\a-1})z_\a +\frac{1}{6}m_\a \nonumber \\
    & & +2q_\a h_{\a+1}^2-2r_{\a}h_{\a-1}^2
\end{eqnarray}
\begin{enumerate}
    \item[(I)] Suppose the shear tensor is degenerate, say $h_1=0$. Then
$h_3=-h_2\neq 0$, $m_1=0$ by $R_1$, while $\{Q_2,R_2,\partial_2 h_1
=0\}$ (resp.~$\{Q_3,R_3,\partial_3(h_1)=0\}$) can be solved for
$H_{13},z_2,q_2$ (resp.~$H_{12},z_3,r_3$) to give:
\begin{eqnarray}
  \label{M13} H_{13} &=& -\frac{m_2}{3h_2},\quad z_2 = \frac{m_2(2\t-h_2)}{6\rho},\\
  q_2&=&\frac{m_2(2h_2\t-h_2^2+2\rho)}{6\rho h_2^2},\nonumber \\
  \label{M12} H_{12} &=& \frac{m_3}{3h_2},\quad z_3 =
  \frac{m_3(2\t-h_2)}{6\rho},\\
  r_3&=& -\frac{m_3(2h_2\t-h_2^2+2\rho)}{6\rho h_2^2}. \nonumber
\end{eqnarray}
Computing the time evolution of $R_2$, resp.~$R_3$ and inserting (\ref{M13}), resp.~(\ref{M12}), one
gets
\begin{eqnarray}
    m_2\,O_1=m_3\,O_1=0
\end{eqnarray}
with
\begin{eqnarray}
O_1\equiv-2\dot{p}h_2+4\dot{p}\t+12\rho^2+4\rho h_2\t+\rho
h_2^2+4\rho p. \end{eqnarray}
 Suppose now $m_2\neq 0$, then $O_1=0$. Applying twice
$\partial_2$ to this equation, in each step making use of
(\ref{eah2}) and $\partial_2 p=\partial_2\dot{p}=0$ (the latter
following from $[\partial_0 , \partial_{\a}]p$,
substituting (\ref{M12}) and
dividing by $m_2$, one gets a linear system $O_1=O_2=O_3=0$ in the
two variables $p,\dot{p}$, parametrized by $\rho,\t,h_2$; the
vanishing of the extended system determinant yields
\begin{eqnarray*}
    P_1&\equiv&  -9h_2^6+54\t h_2^5+274h_2^4\rho-108\t^2h_2^4-648h_2^3\rho\t\\
    & & +72\t^3h_2^3+168h_2^2\t^2\rho +660\rho^2h_2^2+64\rho\t^3h_2\\
    & & -552\t\rho^2h_2-360\rho^3+192\rho^2\t^2=0.
\end{eqnarray*}
Repeating the above process for $P_1$ instead of $O_1$,
one obtains two new polynomial equations  $P_2,P_3$. Eliminating $\t$
between $P_1$ and $P_2$, resp.~$P_1$ and $P_3$ by calculating the
resultant w.r.t.~$\theta$ yields two equations
$h_2^7\rho^8K_1(\rho,\theta)=0$ and $h_2^7\rho^8K_2(\rho,\theta)=0$,
where the resultant of $K_1$ and $K_2$ w.r.t.~$\rho$ is proportional
to $h_2^{80}$.
Hence one obtains the contradiction $\rho=0$ or $h_2=0$ (lemma 4.1 (b)). Thus
$m_2=0$. The same reasoning, involving the same polynomials, holds
when applying $\partial_3$ instead of $\partial_2$ in the above,
such that also $m_3=0$. Hence for degenerate shear, the fluid has a
vanishing spatial gradient of the energy density.

\item[(II)] Suppose the shear is non-degenerate, i.e.~$h_1h_2h_3\neq
0$.
On making use of (\ref{eah1}) and (\ref{eah2}) one can
solve 
$\{[\partial_2 , \partial_3]\rho, [\partial_2 , \partial_3]\t,
\partial_{2}Q_{3},\partial_{3}Q_{2}\}$
for
$\partial_{2}m_{3},\partial_{3}m_{2},\partial_{2}z_{3},\partial_{3}z_{2}$,
yielding
\begin{eqnarray}\label{eam} \partial_{2}m_{3}&=&q_{3}m_{2}-\frac{h_{3}}{h_1}n_{11}m_1\nonumber \\
&& +\left(\frac{3}{2\rho}+\frac{1}{3h_{2}h_{3}}\right)m_{2}m_{3}.
\end{eqnarray}
Then $[\partial_0 , \partial_2]m_{3}$ implies
\begin{eqnarray}\label{Pa}
    && X_1 \equiv 9h_1h_2h_3m_{2}m_{3}\dot{p} + [3h_{2}h_{3}h_1^2(r_{2}m_{3}+q_{3}m_{2})\nonumber \\
    & & +3h_{2}^2h_{3}^2n_{11}m_1-h_1(h_{2}-h_{3})m_{2}m_{3}]\rho^2=0
\end{eqnarray}
where in (\ref{eam}) and (\ref{Pa}) we have eliminated the
$H_{\a+1,\a-1}$ by $Q_\a$ terms and the $z_\a$ by $R_\a$. By
repeating the above for the index couples 31 and 12 instead of 23,
one gets equations $X_2,X_3$, which can formally be obtained by
cyclic permutation of the indices of $X_1$ (twice). Taking the
combination $m_2X_2-m_3X_3$
one gets (with $\rho\neq 0$):
\begin{eqnarray}
   A_0 &\equiv&   3h_1^2h_{3}^2m_{2}^2n_{22}-3h_1^2h_{2}^2m_{3}^2n_{33}\nonumber \\
   & & +3h_1h_2h_3(m_2m_3(h_2q_1-h_3r_1)\nonumber \\
   & & +m_1m_2h_2r_3-m_1m_3h_3q_2)\nonumber\\
   &=& m_1m_2m_3(h_1^2+2h_2h_3).
\end{eqnarray}
On examination of (\ref{e0qr}),
(\ref{e0nn}), (\ref{e0hu}) and (\ref{e0m}) one sees that
calculation of $A_{i+1}\equiv \partial_0 A_i+\frac{(19+2i)}{3}\t
A_i,i=0,\ldots,5$, yields a system $\{A_i,i=0\,..\,6\}$ of 7
linear equations
in
$n_{22},n_{33},q_1,r_1,q_2,r_3$, parametrized by the $m_\a$ and $h_\a$, which can only be consistent if the
determinant of the extended system matrix vanishes. 
This yields $m_1^3m_2^6m_3^6h_1^{17}h_2^{12}h_3^{12}C_1(h_2,h_3)=0$,
where $C_1(h_2,h_3)$ is a homogeneous polynomial of degree 8; thus
if $C_1(h_2,h_3)$ vanished the ratio $h_2/ h_3 $ would be constant,
in contradiction with lemma 1 (c) and the assumed non-degeneracy of
the shear.
Therefore $m_1m_2m_3=0$. As a vanishing spatial gradient of the energy density
(all $m_i$ zero) implies degenerate shear (cf.\ section 4), we are left with two
qualitatively different subcases: (a) $m_1=0,m_2m_3\neq0$ and (b)
$m_2=m_3=0,m_1\neq 0$. In general, if $m_\a=0$ for \emph{fixed} $\a$, it
follows that $H_{\a+1\,\a-1}=z_\a=0$ from (\ref{Qa}) and (\ref{Ra}).
 From (\ref{e0z}), (\ref{e0qr})
and (\ref{e0hu}) one then deduces
$L_\a\equiv h_{\a+1}^2 q_\a-h_{\a-1}^2 r_\a=0$,
$\partial_0 L_\a \equiv
(4h_{\a+1}+5h_{\a-1})h_{\a+1}^2q_\a+(4h_{\a-1}+5h_{\a+1})h_{\a-1}^2r_\a=0$,
which for non-degenerate shear can only be consistent if
$q_\a=r_\a=0$.
\begin{enumerate}
 \item[(a)] $m_1=0,m_2m_3\neq0$: \\
 As also $q_1=r_1=0$, one deduces $n_{22}=n_{33}=0$ from $X_2=X_3=0$, i.e., $n_2=n_3=-n_1=\frac{1}{2}n_{11}$. Then e.g. the (31) component of the field equations together with $\begin{pmatrix} &2&\\1&2&3\end{pmatrix}$ and the (31) component of (\ref{dotsigma}) together with the (2) component of (\ref{divH})
respectively give
\begin{eqnarray}\label{e2n1}
    \partial_{2}n_{1} &=& -2q_2n_{1},\nonumber \\
     \partial_{2}H_{33} &=& (H_{33}-H_{22})r_2-3H_{13}n_{1}.
\end{eqnarray}
while (\ref{Haa}) further reduces to
\begin{eqnarray}\label{Haam1}
H_{22}=n_1h_2,\quad H_{33}=-n_1h_3.
\end{eqnarray}
Propagation of the second equation of (\ref{Haam1}) along the
$\partial_2$ integral curves, using (\ref{eah1}),
(\ref{e2n1}) and (\ref{Haam1}), yields $n_1(4H_{13}+z_2+4h_3q_2)=0$.
As $n_1=0$ is not allowed by (\ref{Haam1}), it follows that
$B_0\equiv 4H_{13}+z_2+4h_3q_2=0$. Calculation of
$B_{i+1}=\partial_0B_i+\frac{3+2i}{3}\t B_i$, $i=0,1,2$ gives a
homogeneous linear system $\{Q_2,B_0,B_1,B_2,B_3\}$ in the
variables $q_2,r_2,m_2,z_2,M_{13}$ and parametrized by  $h_2,h_3$,
which can only be consistent with $m_2\neq 0$ if the determinant,
namely $\frac{3}{2}h_1^2h_2^2h_3^2(21h_2^2+64h_3h_2+16h_3^2)$,
vanishes, which is contradictory to lemma 4.1 (c).
\item[(b)] $m_2=m_3=0,m_1\neq 0$:\\
 Apart from $q_2=r_2=q_3=r_3=0$, one has now $n_{11}=0$ from (\ref{Pa}).
One can solve the (23) component of the field equations together with its $\partial_0$ derivative and
$[\partial_0, \partial_1](n_{22}-n_{33})$
for
$\partial_1{n_{22}},\partial_1 n_{33}$ to give
\begin{eqnarray}
    \partial_1 n_{22} &=& -\frac{1}{2 h_1} \left[(4H_{23}-z_1+4h_2r_1)n_{22}\right. \nonumber \\
    & & +\left. (4H_{23}+z_1+4h_2q_1)n_{33}\right],\\
    \partial_1 n_{33} &=& -\frac{1}{2 h_1} \left[(4H_{23}-z_1-4h_3r_1)n_{22}\right. \nonumber \\
    & & +\left.(4H_{23}+z_1-4h_3q_1)n_{33}\right].
\end{eqnarray}
Then $[\partial_0, \partial_1](n_{22}+n_{33})$,
eliminating $H_{23}$ by $R_1$, gives
\begin{eqnarray}
    C_0 &\equiv& 12\left[h_2n_{22}-(2h_3+3h_2)n_{33}\right]h_2q_1\nonumber \\
    &&+12\left[h_3n_{33}-(2h_2+3h_3)n_{22}\right]h_3 r_1\nonumber\\
    &&+13(n_{22}-n_{33})m_1-6\left[ (h_2+3h_3)n_{22}\right. \nonumber \\
    && \left.+(h_3+3h_2)n_{33}\right]z_1=0.
\end{eqnarray}
Calculation of $C_{i+1}=\partial_0C_i+\frac{6+2i}{3}\t C_i$,
$i=0,\ldots ,3$, eliminating in each step $H_{23}$ by means of
$R_1$, gives two homogeneous linear systems
$\{C_0,C_1,C_2,C_j\}$, $j=3,4$, in the variables
$q_1,r_1,m_1,z_1$ and parametrized by $h_2,h_3,n_{22},n_{33}$. Again
this can only be consistent with $m_1\neq 0$ if the determinants of
the coefficient matrices vanish. However, on computing the resultant
of these determinants w.r.t.~$n_{22}$, resp. $n_{33}$, one gets
equations
$h_1^{20}h_2^5h_3^5P(h_2,h_3)n_{22}^{16}=h_1^{20}h_2^5h_3^5P(h_2,h_3)n_{33}^{16}=0$,
with $P(h_2,h_3)$ a homogeneous polynomial of its arguments, such
that $n_{22}=n_{33}=0$ by lemma 4.1 (c). Hence $n_1=n_2=n_3=0$, in
contradiction with (\ref{Haa}).
\end{enumerate}
Hence case (II) is not allowed, and this concludes the proof.
\end{enumerate}

\end{document}